\documentclass[11pt]{article}
\usepackage{graphicx,rotating,cite,epsfig,amsmath,amssymb}
\usepackage{times}
\usepackage{xspace}
\usepackage{array}

\newcommand{\dstarpl}{\ensuremath{\mathrm{D}^{\ast+}}\xspace}

\newcommand{\dpl}{\ensuremath{\mathrm{D}^+}\xspace}
\newcommand{\dspl}{\ensuremath{\mathrm{D}_{\mathrm{s}}^+}\xspace}

\newcommand{\dzero}{\ensuremath{\mathrm{D}^{0}}\xspace}
\newcommand{\kzeros}{\ensuremath{\mathrm{K}_{\mathrm s}^{0}}\xspace}

\newcommand{\lbzero}{\ensuremath{\mathrm{\Lambda}^0}\xspace}
\newcommand{\lbcpl}{\ensuremath{\mathrm{\Lambda_{\mathrm{c}}^+}}\xspace}

\newcommand{\lb}{\ensuremath{\mathit{\lambda}}\xspace}

\newcommand{\vcd}{\ensuremath{{V_{\mathrm{cd}}}}\xspace}
\newcommand{\BR}{\ensuremath{\mathit{BR}}\xspace}

\newcommand{\GeV}{\ensuremath{\mbox{GeV}}\xspace}

\newcommand{\GeVc}{\ensuremath{\mbox{GeV}/c}\xspace}

\newcommand{\mm}{\ensuremath{\mbox{mm}}\xspace}
\newcommand{\micron}{\ensuremath{\mu \mbox{m}}\xspace}

\newcommand{\mrad}{\ensuremath{\mbox{mrad}}\xspace}

\newcommand{\numu}{\ensuremath{\nu_\mu}\xspace}
\newcommand{\numubar}{\ensuremath{\overline{\nu}_\mu}\xspace}
\newcommand{\nutau}{\ensuremath{\nu_\tau}\xspace}

 1
 1
 1
 3
 2
 2
 2

\begin{document}
\begin{titlepage}

\title{
 \vskip -1.4cm
 {\normalsize 
 \rightline {CERN-PH-EP-2011-109}
 }
 \vglue 1.8cm
 {
 Measurement of charm production in neutrino charged-current interactions}
 }
 
\author{CHORUS Collaboration}
\maketitle

\begin{abstract}
 The nuclear emulsion target of the CHORUS
 detector was exposed to the wide-band neutrino beam of the CERN SPS of
 27 GeV average neutrino energy from 1994 to 1997. 
 In total about 100000 charged-current
 neutrino interactions with at least one identified muon were located
 in the emulsion target and fully reconstructed, using newly developed
 automated scanning systems. Charmed particles were searched for by a
 program recognizing particle decays. The observation of the decay in
 nuclear emulsion makes it possible to select a sample with very low
 background and minimal kinematical bias.
 2013 charged-current interactions with a charmed hadron candidate in the final
 state were selected and confirmed through visual inspection.  
 The charm production rate
 induced by neutrinos relative to the charged-current cross-section is
 measured to be 
 $\sigma ( \numu N \rightarrow \mu^{-} C X )/\sigma
 (\mathrm{CC}) = (5.75 \, \pm \, 0.32 (\mbox{stat}) \, \pm \, 0.30
 (\mbox{syst}))$\%. 
 The charm production cross-section as a function of the neutrino energy
 is also obtained. 
 The results are in good agreement with previous measurements.
 The charm-quark hadronization produces the following charmed hadrons
 with relative fractions (in \%): 
 $f_{\dzero}= 43.7 \,\pm\,  4.5$, 
 $f_{\lbcpl}= 19.2 \,\pm\, 4.2$, 
 $f_{\dpl}= 25.3 \,\pm\, 4.2$, and
 $f_{\dspl}= 11.8 \,\pm\, 4.7$.

\end{abstract}

\vspace{5.0cm}


\newpage

\begin{center}   
{\Large {CHORUS Collaboration}}
\vspace{0.1cm}

\noindent A.~Kayis-Topaksu, G.~\"{O}neng\"ut\\
{\bf \c{C}ukurova University, Adana, Turkey}

\noindent R.~van Dantzig,  M.~de Jong, R.G.C.~Oldeman$^1$\\
{\bf NIKHEF, Amsterdam, The Netherlands}

\noindent M.~G\"uler, U.~K\"ose, P.~Tolun\\
{\bf  METU, Ankara, Turkey}

\noindent M.G.~Catanesi, M.T.~Muciaccia\\
{\bf Universit\`a di Bari and INFN, Bari, Italy}

\noindent K.~Winter\\
{\bf  Humboldt Universit\"at, Berlin, Germany$^{2}$}

\noindent B.~Van de Vyver$^{3}$, P.~Vilain$^{4}$, G.~Wilquet$^{4}$\\
{\bf Inter-University Institute for High Energies (ULB-VUB) Brussels, Belgium}

\noindent B.~Saitta\\
{\bf Universit\`a di Cagliari and INFN, Cagliari, Italy}

\noindent E.~Di Capua\\
{\bf Universit\`a di Ferrara and INFN, Ferrara, Italy}

\noindent S.~Ogawa, H.~Shibuya \\
{\bf Toho University,  Funabashi, Japan}

\noindent I.R.~Hristova$^5$, T.~Kawamura,
D.~ Kolev$^6$, H.~ Meinhard, J.~Panman, A.~Rozanov$^{7}$,
R.~Tsenov$^{6}$, \mbox{J.W.E.~Uiterwijk}, P.~Zucchelli$^{8}$\\
{\bf CERN, Geneva, Switzerland}

\noindent J.~Goldberg\\
{\bf Technion, Haifa, Israel}

\noindent M.~Chikawa\\
{\bf Kinki University, Higashioaka, Japan}

\noindent J.S.~Song, C.S.~Yoon\\
{\bf Gyeongsang National University,  Jinju, Korea}

\noindent K.~Kodama, N.~Ushida\\
{\bf Aichi University of Education, Kariya, Japan}

\noindent S.~Aoki, T.~Hara\\
 {\bf Kobe University,  Kobe, Japan}

\noindent T.~Delbar,  D.~Favart, G.~Gr\'egoire, S.~ Kalinin, I.~ Makhlioueva\\
{\bf Universit\'e Catholique de Louvain, Louvain-la-Neuve, Belgium} 

\noindent A.~Artamonov, P.~Gorbunov, V.~Khovansky, V.~Shamanov, I.~Tsukerman\\
{\bf Institute for Theoretical and Experimental Physics, Moscow, Russian
Federation}

\noindent N.~Bruski, D.~Frekers\\
{\bf Westf\"alische Wilhelms-Universit\"at, M\"unster, Germany$^{2}$}

\noindent K.~Hoshino, J.~Kawada$^{9}$, M.~Komatsu, M.~Miyanishi, 
M.~Nakamura, T.~Nakano, K.~Narita, K.~Niu, K.~Niwa, 
N.~Nonaka, O.~Sato, T.~Toshito\\
{\bf Nagoya University, Nagoya, Japan}

\noindent S.~Buontempo, A.G.~Cocco, N.~D'Ambrosio,
G.~De Lellis, G.~ De Rosa, F.~Di Capua$^{*}$, G.~Fiorillo, A.~Marotta,
M.~Messina, P.~ Migliozzi, L.~Scotto Lavina, P.~ Strolin, V.~Tioukov\\
{\bf Universit\`a Federico II and INFN, Naples, Italy}

\noindent T.~Okusawa\\
{\bf Osaka City University, Osaka, Japan}

\noindent U.~Dore, P.F.~Loverre, L.~Ludovici, 
G.~Rosa, R.~Santacesaria, A.~Satta, F.R.~Spada\\
{\bf Universit\`a La Sapienza and INFN, Rome, Italy}

\noindent E.~Barbuto, C.~Bozza, G.~Grella, G.~Romano, C.~Sirignano, S.~Sorrentino\\
{\bf  Universit\`a di Salerno and INFN, Salerno, Italy}

\noindent Y.~Sato, I.~Tezuka\\
{\bf Utsunomiya University,  Utsunomiya, Japan}

\end{center}

{\footnotesize
---------

\begin{flushleft}

$^{1}$ {now at Universit\`a di Cagliari and INFN, Cagliari, Italy.}
\newline
$^{2}$ {Supported by the German Bundesministerium f\"ur Bildung und Forschung under contract numbers 05 6BU11P and 05
7MS12P.}
\newline
$^{3}$ {Fonds voor Wetenschappelijk Onderzoek, Belgium.}
\newline
$^{4}$ {Fonds National de la Recherche Scientifique, Belgium.}
\newline
$^{5}$ {Now at DESY, Hamburg.}
\newline
$^{6}$ {On leave of absence and at St. Kliment Ohridski University of Sofia, Bulgaria.}
\newline
$^{7}$ {Now at CPPM CNRS-IN2P3, Marseille, France.}
\newline
$^{8}$ {On leave of absence from INFN, Ferrara, Italy.}
\newline
$^{9}$ {Now at University of Bern, Switzerland.}
\newline
$^{*}$ {Corresponding author, email:dicapua@na.infn.it}
\end{flushleft}
}


\end{titlepage}


\newpage

\section {Physics motivation}

About forty years after the discovery of the charm quark at
SLAC~\cite{jpsislac} and BNL~\cite{jpsibnl}, and the first observation 
of charm decay in nuclear emulsion~\cite{niucharm}, the study of the
charmed  particles is still a challenging field of particle physics. 
In particular, the neutrino induced charm-production offers the
possibility to study the strange-quark content of the nucleon, to
measure ``directly'' the CKM matrix element \vcd and to test
models for charm-production and subsequent hadronization. 
Moreover, neutrinos produce charmed hadrons also via
specific processes like quasi-elastic and diffractive scattering which
provide a unique tool for studies of exclusive charm production.

In addition to its intrinsic interest, an improved knowledge of charm
production helps to better understand the charm background in neutrino
oscillation experiments where the signal is given by the
production of a $\tau$ lepton or of muons of apparently ``wrong'' charge
with respect to that expected from the neutrino beam helicity, as in
ongoing experiments~\cite{opera} and at future neutrino
facilities~\cite{nufact}. 

 Charm production in neutrino and anti-neutrino charged-current (CC)
interactions has been studied by several experiments by looking at the
presence of two oppositely charged leptons in the final state. In
particular, CDHS~\cite{CDHS}, CCFR~\cite{CCFR}, CHARM~\cite{CHARM},
CHARM-II~\cite{CHARMII}, NuTeV~\cite{NuTeV} and
CHORUS (using only its electronic detectors)~\cite{CHORUS_dimuon} have
collected large statistics of opposite-sign dimuon events. 
The leading muon is interpreted as originating from the
neutrino vertex and the other one, of opposite charge, as being the
decay product of the charmed particle. Although massive electronic
detectors allow obtaining large statistics, they have some drawbacks.
Of the charmed parent only the decay muon is seen, resulting in an
event sample composed of a mixture of all charmed-particle species
weighted by their muonic branching ratios. Furthermore, experiments of
this type suffer from significant background ($\sim 20 \%$) due to the
undetected decay-in-flight of a pion or a kaon.  
The identification of the primary muon and the decay muon is not
unambigous.  
Moreover the kinematic cuts on the energies of the primary and decay
muons, required for background reduction, make it difficult to study 
cross-sections at energies below 20--30~\GeV.

Unlike dimuon experiments,  BEBC~\cite{BEBC} and NOMAD~\cite{NOMAD} were
able to recognize specific charm decay modes by reconstructing an
invariant mass from the decay daughters. 
Only a few specific decay modes were selected and thus also a specific
parent particle type. 
CHORUS~\cite{dstar} took advantage of the spatial resolution of nuclear
emulsion to distinguish the charm decay vertex from the primary neutrino
interaction vertex. 
In combination with a measurement of the transverse momentum of one of
the decay products it could select a specific decay mode of the
\dstarpl with very low background. 

The use of a hybrid nuclear emulsion
detector was pioneered by the E531~\cite{E531} experiment at FNAL. 
In nuclear emulsion, the different charmed particles are recognized on
the basis of their decay topology and short flight length, so 
that the required kinematic cuts can be quite loose. 
All decay channels are therefore observed, not only the muonic ones,
without requiring knowledge of muonic branching ratios and with very low 
background. 
The disadvantage of the low statistics generally obtained in emulsion
experiments (122 charm events observed in E531) has been overcome in
the CHORUS experiment~\cite{detpap} by using a massive (770 kg)
nuclear emulsion target and automated emulsion
scanning~\cite{aoki,TrackSelector}. A high-statistics sample of charm
decays in emulsion, more than one order of magnitude larger than in
E531, has thus been collected as reported in this paper.

The CHORUS experiment took data from 1994 to 1997 in the CERN Wide
Band Neutrino Beam~\cite{beam} which essentially consisted of muon
neutrinos. 
The analysis presented
here is based on the complete CHORUS sample of 2013 charm events,
confirmed by visual inspection. The visual inspection recognized 1048
events as due to the production of the neutral charmed hadron \dzero
and 965 events as due to the production of a charged charmed hadron
\lbcpl, \dpl or \dspl. The analysis of the \dzero events has
been reported in a previous publication~\cite{dzero}. The charged
sample is analysed in this paper. The relative contribution of the
different charmed hadrons to the total charged sample is obtained from
a likelihood approach by using the decay lifetime information. 

The neutral and charged charm production candidates are combined for the 
measurement of the 
total charm production rate relative to charged-current neutrino
events averaged on the neutrino energy spectrum as well as of its
dependence on neutrino energy.

\section {The experimental set-up}

The CHORUS experiment was designed to investigate
neutrino oscillation by searching for the \nutau appearance in
the SPS wide-band neutrino beam at CERN through the direct observation
of the $\tau$ decay in nuclear emulsions.  Since charmed particles
have a flight length comparable to that of the $\tau$ lepton, the
experiment is also suitable for the study of charm production. The
detector, described in more detail in \cite{detpap}, uses a hybrid
approach that combines a nuclear emulsion target with electronic
detectors.

The emulsion target, of 770~kg total mass, is segmented along the beam
direction in four stacks of $1.4 \times 1.4$~m$^{2}$ transverse area
and about 3~cm thickness. It is equipped with high-resolution trackers
made out of three interface emulsion sheets and a set of scintillating
fibre tracker planes which provides predictions of particle
trajectories into the emulsion stack with an accuracy of about
150~\micron in position and 2~\mrad in angle.

The emulsion scanning is performed by computer-controlled, fully
automated microscope stages equipped with a CCD camera and a read-out
system called `track selector'~\cite{aoki,TrackSelector}.  The
track-finding efficiency is higher than 98\% for track slopes up to
400~\mrad. 

The electronic detectors downstream of the emulsion target include a
hadron spectrometer, a calorimeter and a muon spectrometer.  The
hadron spectrometer measures the bending of charged particles inside
an air-core magnet. The calorimeter is used to determine the energy
and direction of showers. The muon spectrometer provides the charge
and momentum of muons and provides a rough measurement of the leakage
of hadronic showers out of the calorimeter. Several planes of
scintillator hodoscopes are used for triggering the data acquisition
system \cite{det_trig}.

\section{Data collection}
\label{sec:emulsion}

The CHORUS detector was exposed to the wide-band neutrino beam of the
CERN SPS during the years 1994--97, with an integrated flux of $5.06
\, \times \, 10^{19}$ protons on target. The beam, of 27~GeV average
energy, consists mainly of \numu's with a 5\% \numubar
component of 18~GeV average energy.

The series of steps in the location process of a CC event starts with
the track reconstruction in the electronic detectors, including the
muon identification and terminates with the association to the primary
and possibly secondary vertices of the tracks recorded in a volume of
emulsion. The event location process is summarized in \cite{dzero} and
detailed in \cite{murat} and \cite{bart}. The so-called `NetScan'
method used to analyse the emulsion volume around the interaction
point is described in \cite{murat} and \cite{NetScan}.

About 150~000 events have been located in the emulsion target and have
been analysed following this procedure.

An event is recognized as a charged-current neutrino interaction if
the primary muon track, defined by the electronic detectors, is found
in more than one emulsion plate. Decay topologies are selected using
the following criteria.  At least one of the tracks connected to a
secondary vertex is detected in more than one plate, and the direction
measured in the emulsion matches that of a track reconstructed in the
fibre tracker system.  The parent angle is within 400~\mrad
from the beam direction. In the case of a neutral particle decay, the
parent angle is deduced from the line connecting the primary and
secondary vertex.  The impact parameter to the primary vertex of at
least one of the daughter tracks is larger than a value which is
determined on the basis of the resolution~\footnote{The resolution to
extrapolate to the vertex depends on the track angle $\theta$ with
respect to the beam according to the relation $\sigma =
\sqrt{\mbox{0.003}^2+(\mbox{0.0194}\cdot \tan{\theta})^2}\ \mm$.}.
To remove random association, the impact parameter is also
required to be smaller than a value depending on the distance over
which the track is extrapolated to the vertex, typically of the order
of 130~\micron. The flight length of the parent candidate is
required to be larger than 25~\micron.

\begin{table}[tb]
\caption{Charged-current data sample and charm candidates.}
\label{tab:sample}
\vspace{0.5\baselineskip}
\begin{center}
\begin{tabular}{l r}
\hline
Located CC events & \mbox{93807} \\
Selected for visual inspection & \mbox{2752} \\
\hline
Decay topologies with flight length $<$ 25 \micron & 3 \\
Topologies with kink angle $<$ 50~\mrad & 11 \\
Secondary interactions & 278 \\
Electron--positron pairs & 95 \\
Overlaid neutrino interactions & 44 \\  
Uncorrelated (overlaid) secondary vertices & 21 \\  
Passing-through tracks & 128 \\
All tracks from primary vertex & 142 \\
$\delta$-rays & 2 \\
Other & 15 \\             
\hline
Charged charm candidates & 965 \\  
C1 & 452 \\
C3 & 491 \\
C5 & 22 \\
Neutral charm candidates & 1048 \\ 
V2 & 819 \\
V4 & 226 \\
V6 & 3 \\
\hline
Total charm candidates &  2013\\
\hline
\end{tabular}
\end{center}
\end{table}

Out of a sample of \mbox{143742} located neutrino-induced
charged-current interaction vertices, \mbox{93807} were fully scanned
and analysed.
The selection criteria retain \mbox{2752} events as having a decay
topology. 
These have all been visually inspected. 
The presence of a decay was confirmed for 2013 events.
A secondary vertex is accepted as a decay if the
number of charged particles is consistent with charge conservation and
no other activity (Auger-electron or visible recoil) is observed.
The purity of the automatic selection is 73.2\%.

The result of the visual inspection is given in Table~\ref{tab:sample}
where according to the prong multiplicity the observable decay
topologies are classified as even-prong decays V2, V4 or V6 for
neutral particles (mainly \dzero) and odd-prong decays C1, C3 or C5
for charged particles (mainly \lbcpl, \dpl, \dspl).
The rejected sample consists of secondary hadronic interactions,
$\delta$-rays or gamma conversions, overlaid neutrino interactions, and
of low-momentum tracks which, because of multiple scattering appear as
tracks with a large impact parameter.  The remainder consists either
of fake vertices, being reconstructed using one or more background
tracks, or of vertices with a parent track not connected to the
primary (passing-through tracks not identified as such because of
inefficiencies).

As shown in Table~\ref{tab:sample} we find 965 charged charm
candidates (452 with C1 topology, 491 with C3 and 22 with C5) and 1048
neutral charm candidates (819 with V2 topology, 226 with V4 and 3 with
V6).

\section{Reconstruction efficiency and background evaluation}
\label{effbg}

The efficiency of the event reconstruction in the electronic detector
as well as those of the event location and reconstruction in the emulsion,
need to be evaluated.

When the neutrino scatters off a nucleon, several
physical mechanisms produce charmed hadrons. However, they are
predominantly produced in deep-inelastic interactions. 
Different Monte Carlo generators are used~\cite{ilyaproceedings}.
The neutrino beam spectrum is simulated using the GBEAM~\cite{gbeam}
generator based on GEANT3~\cite{geant}. It uses
FLUKA98~\cite{fluka98} to describe the interactions of protons with
the beryllium target.

Deep-inelastic scattering interactions are simulated using the JETTA
generator~\cite{jetta} which is based on LEPTO~6.1~\cite{lepto} and
JETSET~\cite{jetset}.  This generator is used to simulate
charm-production as well as inclusive CC interactions.  Quasi-elastic
interactions and resonance production processes are simulated with the
RESQUE generator~\cite{resque}. In addition, some other
charm-production mechanisms are simulated: quasi-elastic charmed
baryon production by QEGEN~\cite{qegen} and diffractive production of
charmed mesons by the \mbox{ASTRA} generator~\cite{astra}.

The simulation of the detector response as well as the performance of
the pattern recognition in the electronic tracking detectors is
performed for each process by a GEANT3~\cite{geant} based simulation
program.  
The simulated response of the electronic detectors is processed through 
the same analysis chain as the raw data obtained with the detector.
The event location technique in emulsion is parametrized by a
function of the primary muon momentum and angle, taking into account
that the muon momentum distribution is different for the two samples
of CC events containing charm or not.

To evaluate the efficiency to reconstruct decay topologies
of the charmed hadrons, realistic conditions of track
densities in the emulsion have to be reproduced.  These are obtained
by merging the emulsion data of simulated events with real NetScan
data which do not have a reconstructed vertex but contain tracks which
stop or pass through the NetScan fiducial volume.  These so-called
`empty volumes' represent a realistic background.  The combined data
are passed through the same NetScan reconstruction and selection
programs as used for real data.  The details of the response of the
automatic microscopes are used in this calculation.
Important parameters are the angular resolution and efficiency as
function of the incident angle of the track.

To evaluate the detection efficiency for charmed hadrons, the branching
ratios and the corresponding uncertainties are taken into
account. The contribution from QE and DIS interactions to the production
of charmed baryons is evaluated as discussed in \cite{QE}.  The
contribution of diffractive charm production is evaluated by
using the method described in \cite{delellis}. The \dzero
detection  efficiency is given in \cite{dzero}.
Only ratios of the electronic reconstruction and vertex
location efficiencies need to be determined, thus reducing significantly
the systematic error. 
The overall selection efficiencies relative to the selection of CC
events for different decay topologies are shown in
Table~\ref{tab:effch}. 
The requirement that at least one track of the secondary vertex be matched
with a track in the electronic detectors causes the efficiency to be higher
with increasing number of prongs at the decay vertex.

\begin{table*}[tbp] 
\caption{ Overall selection efficiency relative to CC containing
  geometrical acceptance and reconstruction efficiency for charged
  charmed hadrons decaying into one, three and five prongs,
  respectively.}
\label{tab:effch} 
\small 
\vspace{0.5\baselineskip}
\begin{center} 
\begin{tabular}{cccc} 
\hline
 &  $\mathbf \lbcpl$ & $\mathbf \dpl$ & $\mathbf \dspl$\\
\hline
$C^+ \rightarrow \mathrm{1p}$ (\%) & 17.1 $\pm$ 1.3 & 21.7 $\pm$ 0.9 & 23.9 $\pm$ 1.2 \\
\hline
$C^+ \rightarrow \mathrm{3p}$ (\%) & 40.8 $\pm$ 1.6 & 49.0 $\pm$ 1.2 & 57.7 $\pm$ 1.4 \\
\hline
$C^+ \rightarrow \mathrm{5p}$ (\%) & 44.2 $\pm$ 5.2 & 52.7 $\pm$ 6.5 & 57.3 $\pm$ 3.4 \\
\hline
$\epsilon_{\mathrm{3p}}/\epsilon_{\mathrm{1p}}$ & 2.3 $\pm$ 0.2 & 2.3 $\pm$ 0.1 & 2.4 $\pm$ 0.1 \\
\hline
\end{tabular} 
\end{center} 
\end{table*}

\begin{figure}[tb!]
  \begin{center} 
\scalebox{0.40}{\includegraphics{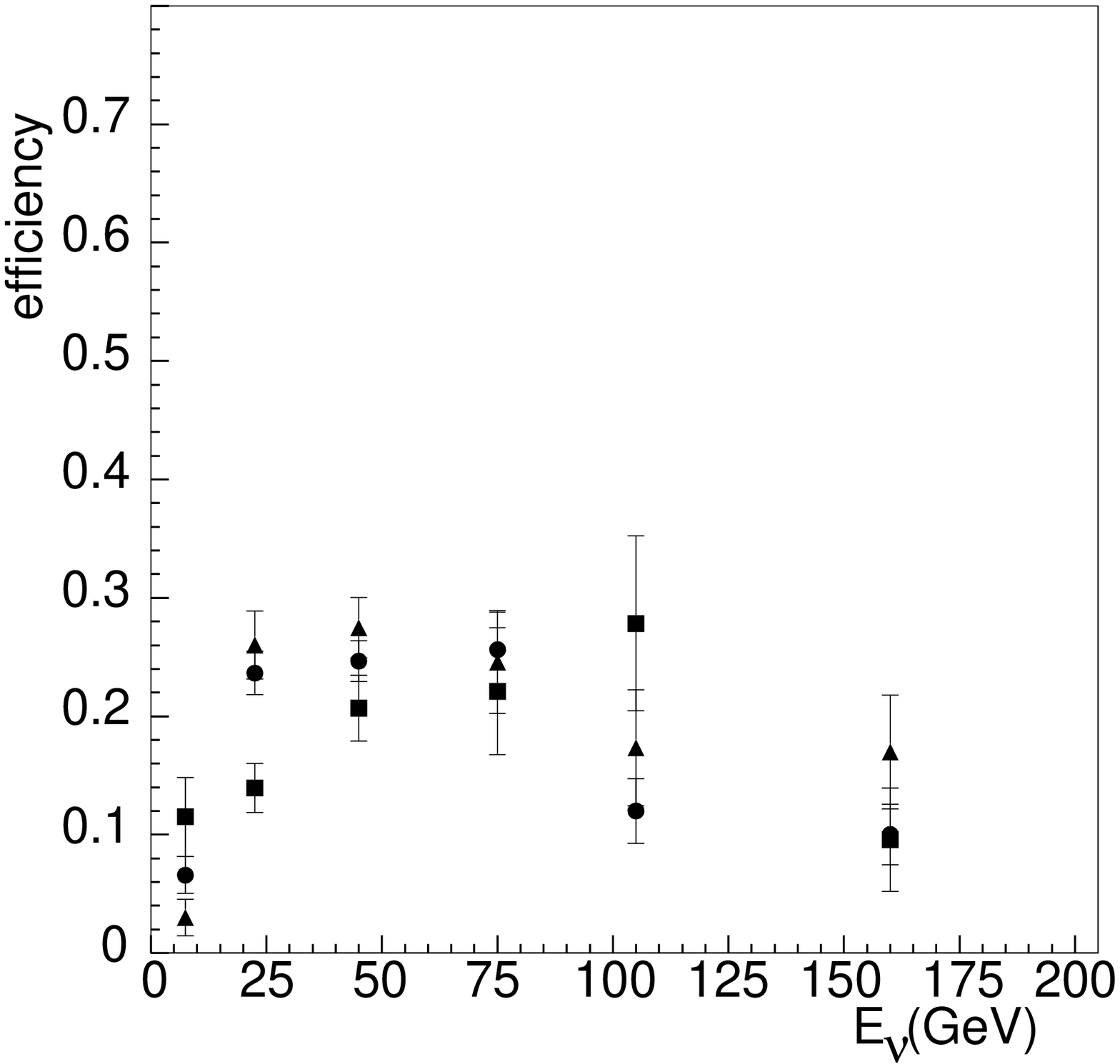}} 
\scalebox{0.40}{\includegraphics{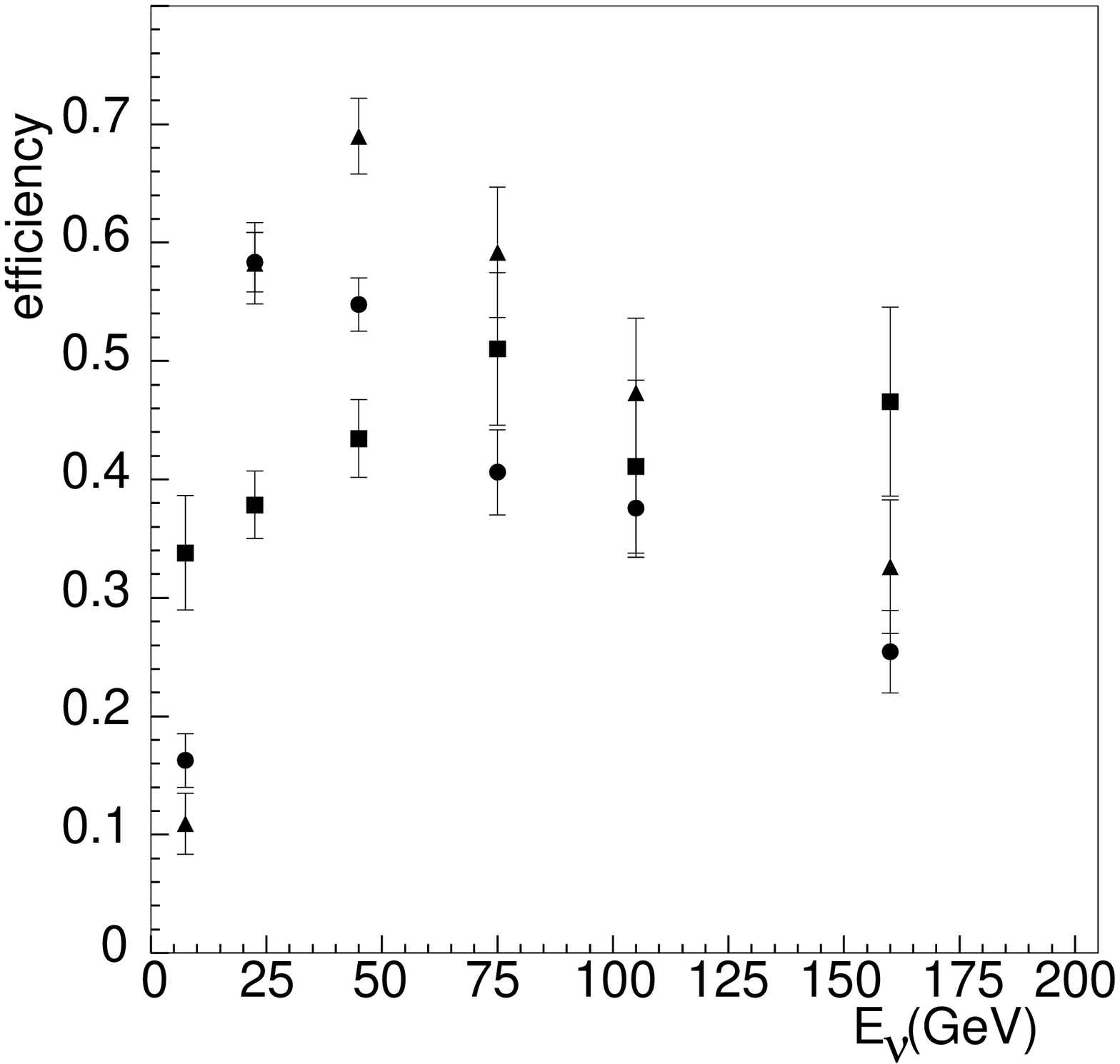}}
  \caption{
  \label{fig:efficiency3p}
   Detection efficiency of charged charm hadrons relative to CC
   interactions as a function of
   neutrino energy for
   1-prong decay (left panel) and 3-prong decays (right panel).  The
   data points indicated with circles show the efficiency for \dpl
   detection, the points marked with triangles are for \dspl detection
   and squares for \lbcpl.  }
\end{center}
\end{figure}

Figure \ref{fig:efficiency3p} shows the detection efficiency of
charged charmed hadrons \dpl, \dspl and \lbcpl relative to CC
interactions as a function of neutrino energy. Two factors make the
selection less efficient at small visible energies: the decay angle
of the charm daughters is larger; the flight length of the charm
parent is shorter and thus a secondary track might be wrongly attached
to the primary vertex. At high energies, a large fraction of charmed
hadron decays near the edge or beyond the fiducial volume.

The spread in the performance of the microscopes is found to induce a
difference of $\pm 2\%$ in the calculation of the selection
efficiencies for the charm detection. The weighted average performance
of the individual microscope stages is taken for the calculation in
order to minimize the uncertainty.  The uncertainty on the efficiency
combination of several charm production mechanisms introduces an
additional error on the efficiency of $\pm 12\%$ for \lbcpl
and $\pm 3\%$ for \dspl.  
Including also other  
factors, such as the uncertainty in the fragmentation, we estimate a
total systematic uncertainty in the efficiency of 14\% for
\lbcpl, 5\% for \dpl and 6\% for \dspl relative to CC
event detection.

There is a small fraction of non-charm events in the manually
confirmed sample. This contamination is mainly due to hadronic
interactions with no heavily ionizing tracks or other evidence for
nuclear break-up (blobs or Auger electrons) that fake charm decays
(white kinks) and decays of $\Sigma^\pm$, \kzeros and
\lbzero.  The backgrounds from the decays of strange
particles were estimated using the JETTA~\cite{jetta} MC generator.

In the \dzero sample, the strange-particle decay background has been
evaluated to be $11.5\,\pm\,1.9$ \lbzero's and $25.1\,\pm\,2.9$
\kzeros's in the V2 sample and negligible for the other \dzero
decay topologies~\cite{dzero}. For charged charmed hadrons, the
expected background in the C1 sample from the decay of charged strange
particles is 8.5$\,\pm\,$1.3 events.

The background due to white kink interactions is obtained by generating
such kind of interactions assuming a hadron interaction length
$\lb = 24$~m~\cite{Satta} and processing them through the full
simulation chain.  The contamination of white kink interactions is
evaluated to be 34.6$\,\pm\,$2.0 in the C1 sample and 3.8$\,\pm\,$0.4 and
1.5$\,\pm\,$0.2 in C3 and C5 samples, respectively.

\section{Charmed particle  production fractions}

\begin{figure}[tb]
  \begin{center} 
   \scalebox{0.4}{\includegraphics{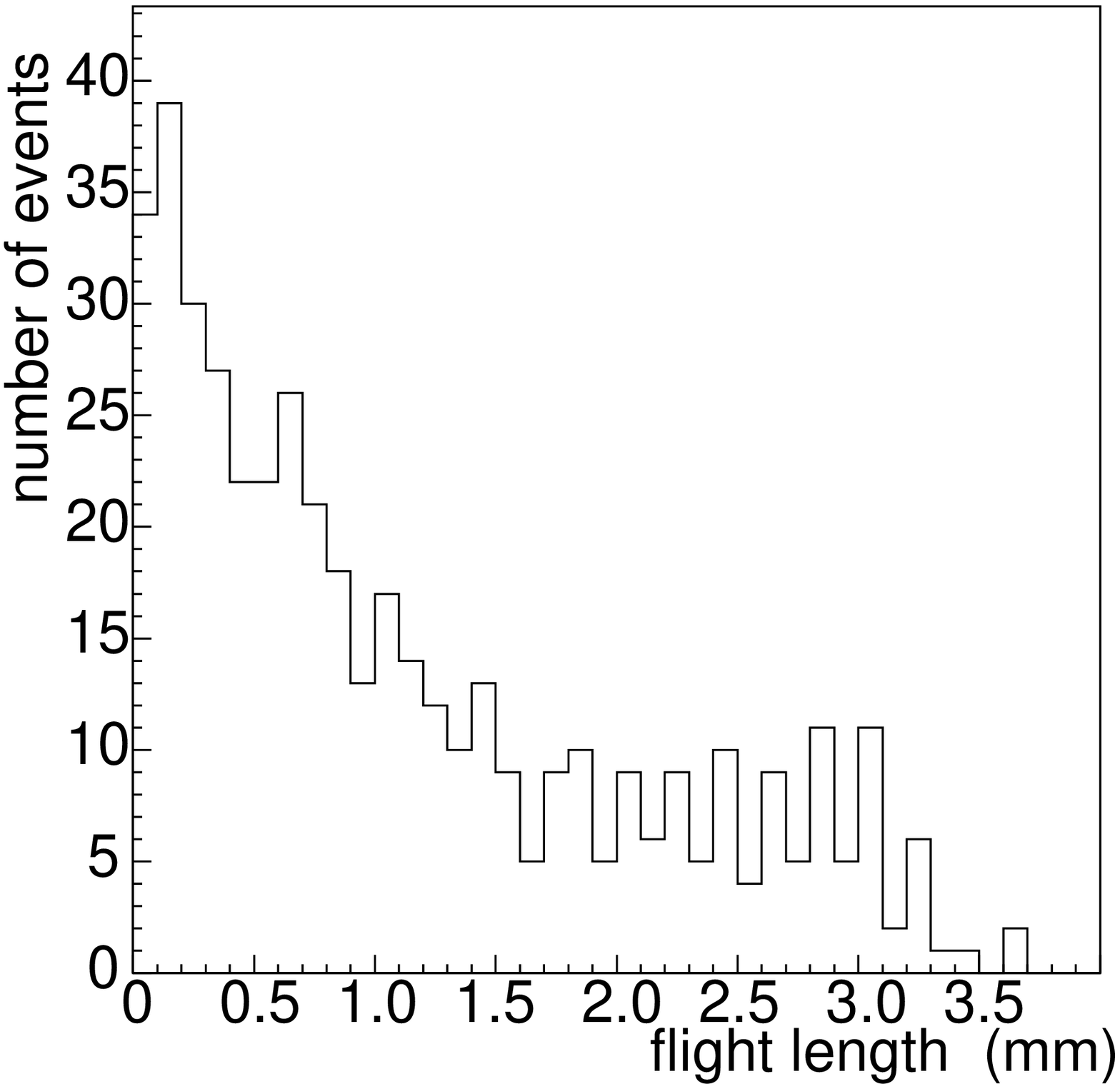}}
  \scalebox{0.4}{\includegraphics{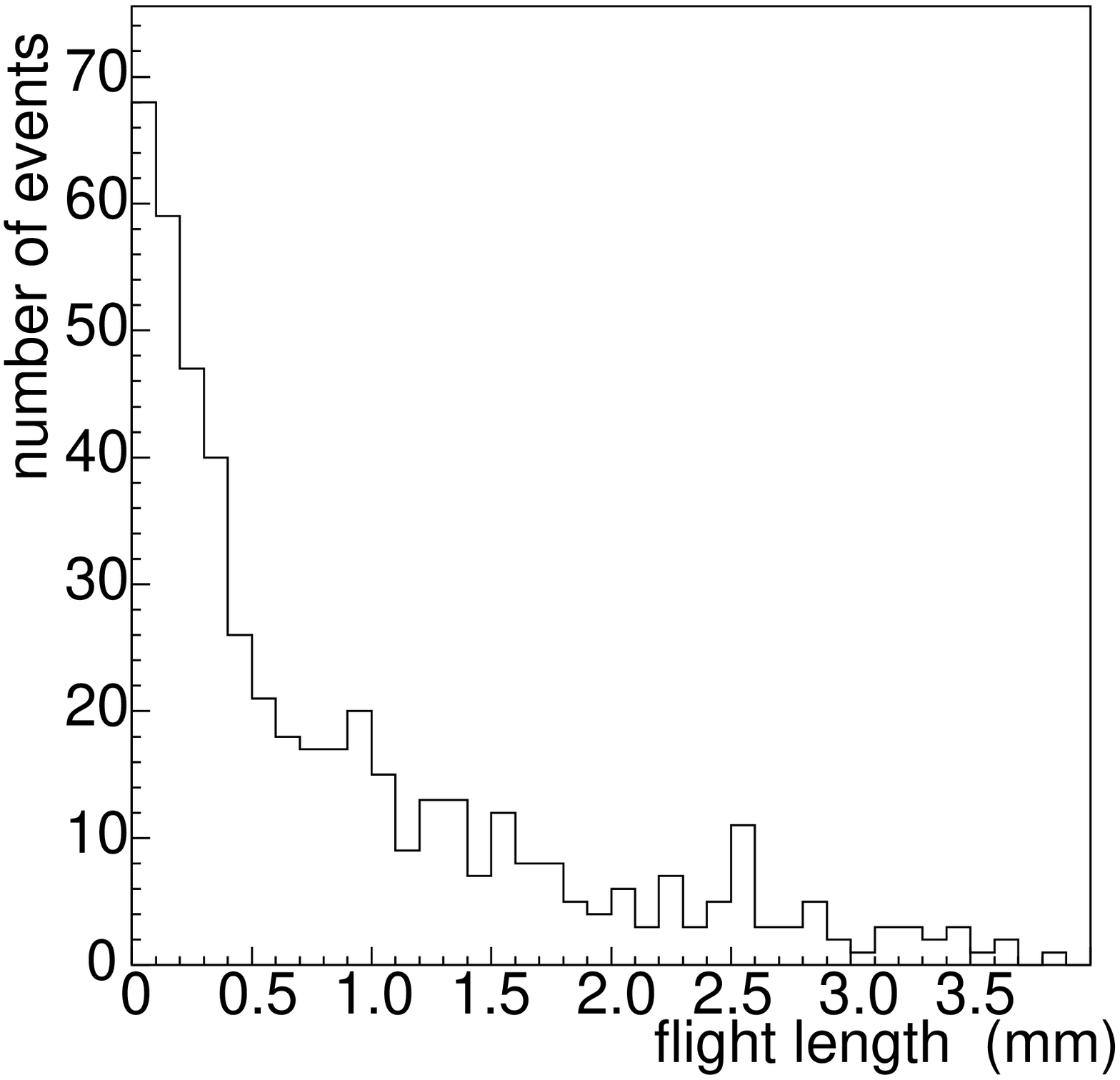}}
  \caption{The flight length distributions for the 1-prong (left) and
   3-prong (right) charged charm events.
   The distributions are truncated due to the limited NetScan volume.
  \label{fig:flight}
  }
\end{center}
\end{figure}

\begin{figure}[tb]
  \begin{center} 
   \scalebox{0.4}{\includegraphics{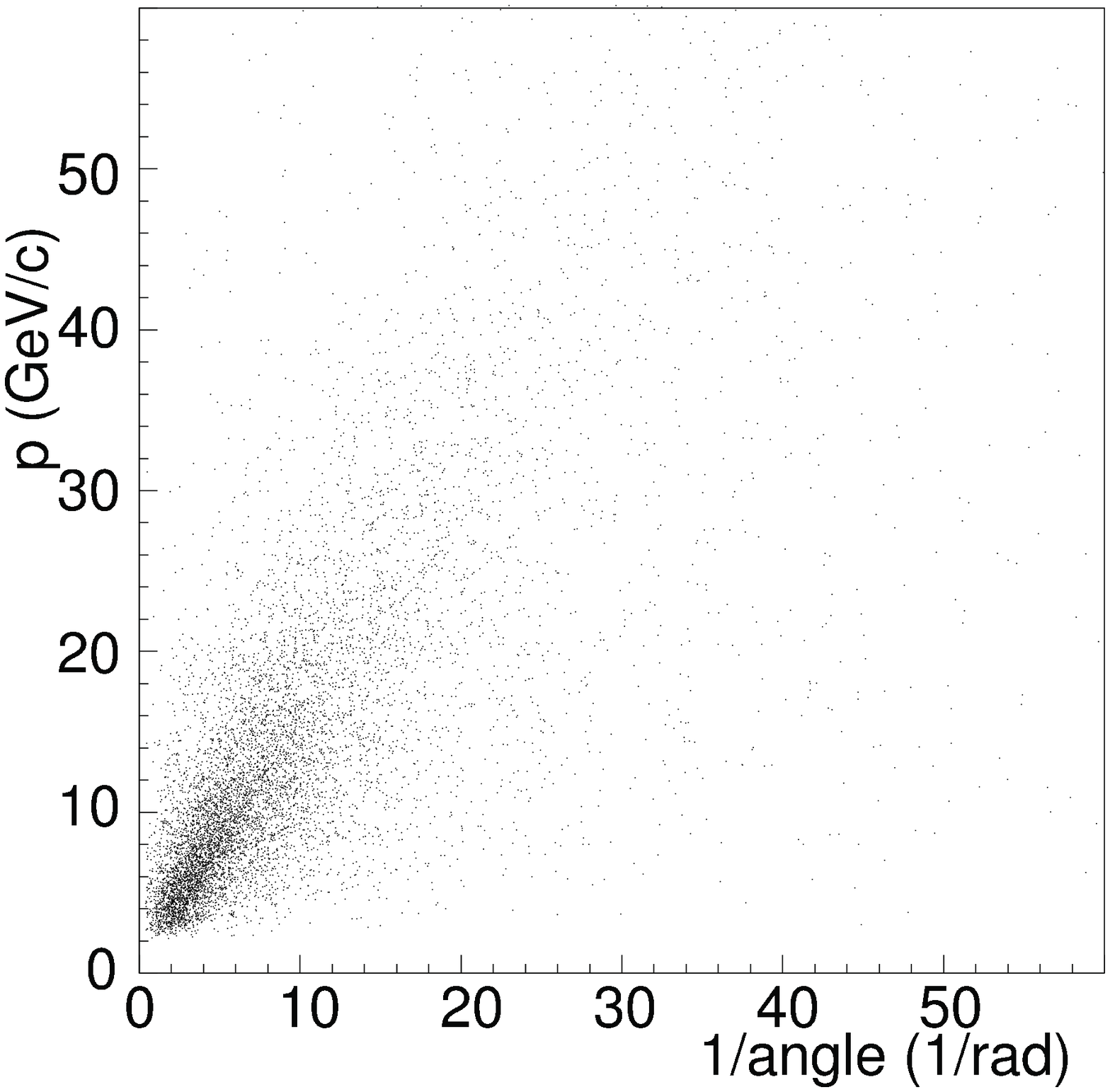}}
  \scalebox{0.4}{\includegraphics{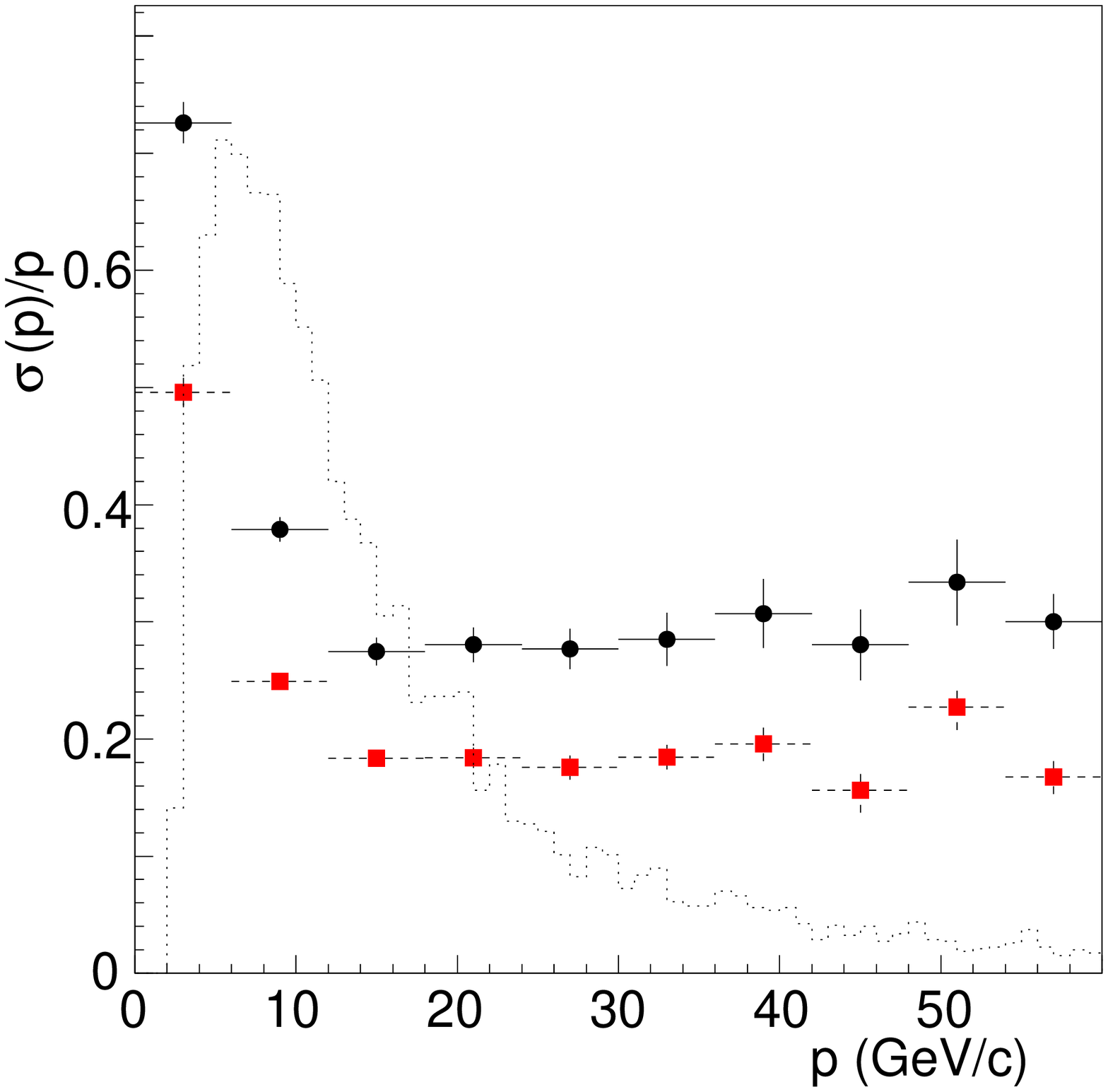}}
  \caption{ Left: correlation between charm momentum and inverse
   daughter's opening angle; Right: charm momentum resolution obtained
   with the Monte Carlo parametrization.  Circles indicate 1-prong
   events, squares 3-prong events. The simulated charm momentum spectrum
   is superimposed in arbitrary units.
  \label{fig:res}
  }
\end{center}
\end{figure}

Since it is not possible to identify the type of charged charmed
particles on an event-by-event basis, they are separated using a
statistical approach by exploiting the different
lifetimes of \lbcpl, \dpl and \dspl, hence by measuring the
flight length and the momentum of the charmed hadrons. The flight length
is very precisely measured in the emulsion target.
The flight length distributions for the 1-prong and 3-prong events are
shown in Fig.~\ref{fig:flight}. 
The momentum is not directly measured, but it can be estimated
exploiting the correlation between the momentum and the decay angle of
the products~\cite{PetreraRomano}. 
For a given decay mode this correlation is determined by the decay
kinematics.  
Figure~\ref{fig:res} shows the correlation between charm momentum and
the daughter's inverse opening angle. 
The charm parent momentum is obtained from the opening angle of the
decay products using a parametrization evaluated with simulated events. 
The resolution obtained with this method is about 25\% for 3 prong
events and 35\% for 1 prong events.

To achieve statistical separation of the
different charged charm species, a likelihood function is constructed
for each event using the decay lifetime information. Following
Ref.~\cite{Bolton2} the form of the probability function for each
event $(n)$ is expressed as the sum of probabilities for the three final
state particle hypotheses $i$ ($i = \lbcpl, \dpl, \dspl$).
Using the numbers of hadrons of each species, $N_i$, as the free
parameters of the fit, the probability takes the form:
$$
P(n)= \frac{ \sum_i {N_i w_i(n) \epsilon_i [l(n)] 
(\frac{M_i u}{c\tau_ip_i(n)}) e^{-\frac{M_il(n)}{c\tau_ip_i(n)}}
}}{\sum_iN_i} \, ,
$$
where $l(n)$ is the measured decay length and $p_i(n)$ is
and the estimated momentum for the hypothesis $i$.
The efficiencies $\epsilon_i(l)$ are a function of the decay length for
each different hadron species $i$.
The mean lifetimes $\tau_i$ and the masses of the charmed hadrons $M_i$
are taken from~\cite{PDG}. 
The weights 
$$
w_i=\big[ \int{\frac{M_i u}{c\tau_iP_i}e^{\frac{-l
M_i}{c\tau_i P_i}} \epsilon_i(l)} dl \big]^{-1}
$$
account for the lifetime spectrum deformation due to selection
efficiencies. 
We have introduced $u$, an arbitrary unit length.

From the probability functions for each event, an
extended log-likelihood function is constructed:
\begin{displaymath}
L=-\sum_n \log{P_n} -N_{\mathrm{obs}}\log{(N_{\lbcpl} + N_{\dpl} + N_{\dspl} )}
+ (N_{\lbcpl} + N_{\dpl} + N_{\dspl} )
\end{displaymath}
The second and third terms
above are the log of the Poisson probability function to observe
$N_{\mathrm{obs}}$ events given the produced events. The Poisson term
incorporates the finite statistics of the experiment. The negative
log-likelihood function is then minimized.
To be independent of charm topological branching ratios, the one-prong and
three-prong samples are fitted separately. 

For the one-prong sample, out of 93807 CC events, the result of the
fit is
$$N_{\lbcpl}^{\mathrm{1p}}=514\,\pm\, 178 \,\pm\,
72~~~~~N_{\dpl}^{\mathrm{1p}}=980\,\pm\, 192 \,\pm\,
50~~~~~N_{\dspl}^{\mathrm{1p}}=449\,\pm\, 235 \,\pm\, 27 \, ,$$
and for the three-prong sample
$$N_{\lbcpl}^{\mathrm{3p}}=507\,\pm\, 88 \,\pm\, 61~~~~~N_{\dpl}^{\mathrm{3p}}=368\,\pm\, 88 \,\pm\,
15~~~~~N_{\dspl}^{\mathrm{3p}}=173\,\pm\, 102 \,\pm\, 10 \, ,$$ where
the first error is the statistical error given by the fit and the second
is due to the systematic effect on the efficiencies discussed in the
previous section. 
Given the small statistics, the five-prong sample is included as a
correction. 
This approximation has a negligible effect on the final
result owing to the small value of this branching ratio.
The relative contributions of charged charm species are:
$$f_{\lbcpl}=(34.1\,\pm\, 7.8)\%~~~~~f_{\dpl}=(44.9\,\pm\,
8.4)\%~~~~~f_{\dspl}=(21.0\,\pm\, 8.6)\% \, .$$

The correlation coefficients are relatively large and similar for the
one-prong and three-prong fits.
We find $\rho(\lbcpl,\dpl) \approx 0.3$, 
$\rho(\lbcpl,\dspl) \approx -0.65$, and 
$\rho(\dpl,\dspl) \approx -0.75$.

\section{Topological branching ratios}

From the results given in the previous section it is possible to estimate
the inclusive topological decay modes for the different charged charm
species. 
In spite of the relatively large errors this information is useful
given the fact that, for each charged charm species, the existing
measurements cover only half of all decay modes. 
We find:
\begin{eqnarray}
 \BR(\lbcpl \rightarrow 3~\mathrm{prongs})&=&(0.49 \,\pm\,
  0.15)\nonumber \\
 \BR(\dpl \rightarrow 3~\mathrm{prongs})&=&(0.27 \,\pm\, 0.08)\\
 \BR(\dspl \rightarrow 3~\mathrm{prongs})&=&(0.27 \,\pm\, 0.19)
  \nonumber \, .
\end{eqnarray}

The value of the \lbcpl 3-prong branching fraction is 1.5 standard
deviations from the one quoted in a previous CHORUS
publication~\cite{lambdac}.
In the present analysis no assumption is made on the other charmed
hadron topological branching ratios, while in Ref.~\cite{lambdac} a
specific assumption had been made.
It should also be noted that the decay-recognition efficiencies are
significantly different in the analysis of \cite{lambdac} compared to the
present analysis.
Owing to advances in the automatic pattern recognition it is possible to
define larger tolerances on the distance of closest 
approach of the decay daughter with respect to the primary muon.
In addition, in Ref.~\cite{lambdac} an equal fraction of QE to DIS \lbcpl
production was assumed, while in this paper the value of 
$0.15 \, \pm \, 0.09$ obtained in Ref.~\cite{QE} was used.
The samples in the two analyses are largely independent due to the
smaller initial sample available in  Ref.~\cite{lambdac} and the
different cuts applied.

The number of charmed hadrons decaying into 5 charged particles is 22
with a background of 1.5 events. This is too small to fit the
different contributions. Assuming that the 5-prong decays are equally
distributed among the three charged charm species and correcting for
the efficiency we have $N_{C5}=42.6 \,\pm\, 9.1$.
The overall charged charm topological branching fractions are:
\begin{eqnarray}
 \BR(C^+ \rightarrow 1~\mathrm{prongs})&=&(0.64 \,\pm\, 0.10) \nonumber \\
 \BR(C^+ \rightarrow 3~\mathrm{prongs})&=&(0.35 \,\pm\, 0.06) \\
 \BR(C^+ \rightarrow 5~\mathrm{prongs})&=&(0.014 \,\pm\, 0.003) \nonumber
   \, .
\end{eqnarray}

\section{\boldmath$\mathbf \dzero$ production cross-section}

The cross-section for the production of neutral charmed meson \dzero in
neutrino CC interactions has been measured using the same sample of
charm candidates~\cite{dzero}. 
The analysis was based on the sample of \dzero
decaying into four charged particles and on the well measured branching
ratio $\BR(D^0 \rightarrow 4 \ \mathrm{prongs})$.  
By using
the same method with the updated value quoted in~\cite{PDG},
$ \BR(\dzero \rightarrow 4~\mathrm{prongs})=0.143 \,\pm\, 0.005$, we obtain the value of
the cross-section
\begin{equation}
\sigma ( \numu N \rightarrow \mu^{-} \dzero X )/\sigma ( \numu N
\rightarrow \mu^{-} X ) = (2.52 \,\pm\, 0.17 (\mbox{stat}) \,\pm\, 0.12
(\mbox{syst})) \% \, .
\end{equation}

It is important to observe that in Ref.~\cite{dzero} also the decay of
\dzero into a fully neutral final state was indirectly measured by subtracting
the branching fractions for 2, 4 and 6 prongs from unity.  
The updated value is
\begin{equation}
\BR(\dzero \rightarrow 0~\mathrm{prongs}) = 0.17 \,\pm\, 0.06 (\mbox{stat}) \,\pm\, 0.03
(\mbox{syst})  \, .
\end{equation}

The latter measurements together with the topological branching ratios
quoted above have an effect on the determination of
the muonic branching ratio of charmed hadrons as reported in
Ref.~\cite{Bmu2}. The updated value is
\begin{equation}
B_\mu= (8.1\,\pm\, 0.9 (\mbox{stat}) \,\pm\, 0.2 (\mbox{syst})) \% \, .
\end{equation}
This is in good agreement with value of $B_\mu= (9.6\,\pm\, 0.4
(\mbox{stat}) \,\pm\, 0.8 (\mbox{syst})) \%$ obtained in the CHORUS dimuon
event analysis~\cite{CHORUS_dimuon}.

\section{Charm cross-sections}

By using the fitted quantities of the one prong and three prong
samples and the corrected number of 5-prong events, a relative
cross-section
\begin{equation}
\sigma ( \numu N \rightarrow \mu^{-} C^+ X )/\sigma ( \numu N
\rightarrow \mu^{-} X ) = (3.23 \,\pm\, 0.27(\mbox{stat}) \,\pm\, 0.21
(\mbox{syst})) \%
\end{equation}
\noindent
is obtained for the charged charm production rate in charged-current
interactions.
Forcing the \lbcpl 1-prong to 3-prong ratio to be that of
Ref.~\cite{lambdac} hardly affects the total charm cross-section (by
about one-quarter of the systematic error).

Including the result obtained for the neutral charmed meson \dzero
given in the previous section, the relative inclusive charm production
rate in charged-current interactions is
\begin{equation}
\sigma ( \numu N \rightarrow \mu^{-} C X )/\sigma ( \numu N
\rightarrow \mu^{-} X ) = (5.75 \,\pm\, 0.32(\mbox{stat}) \,\pm\, 0.30
(\mbox{syst})) \%
\label{eq:charmcross}
\end{equation}
with a relative contribution of the charm species:
$$f_{\dzero}=(43.7\,\pm\, 4.5)\%~~~~~f_{\lbcpl}=(19.2\,\pm\,
4.2)\%~~~~~f_{\dpl}=(25.3\,\pm\, 4.4)\%~~~~~f_{\dspl}=(11.8\,\pm\,
4.7)\% \, .$$

In Ref.~\cite{AntiCharm} we reported 
that in anti-neutrino CC interactions 
$\sigma ( \numubar N \rightarrow \mu^{+} \bar{C} X
)/\sigma ( \numubar N \rightarrow \mu^{+} X ) =
(5.0^{+1.4}_{-0.9}(\mbox{stat}) \,\pm\, 0.7 (\mbox{syst})) \%$. 
The value is similar to what we find for neutrino interactions as
expected since both total CC cross-section and charm production are
about half in this case.

 The energy dependence of the relative charm production cross-section
is obtained by estimating the energy of the interacting neutrino on
an event-by-event basis.  A good estimate is the sum of the energy of
the primary muon and the total energy deposition in the calorimeter
corrected for the energy deposited by the muon and for the unmeasured
energy loss of hadrons in the material upstream of the calorimeter.
The unmeasured part is mainly due to the absorption in the emulsion
stacks and corrected to the measured vertex position.  The resolution
of the calorimeter energy measurement is $ \sigma(E)/E = (0.323 \,\pm\,
0.024)/\sqrt{E/\GeV}+(0.014 \,\pm\, 0.007)\ $~\cite{detpap}. The momentum
resolution varies from $\sim$15\%~\cite{detpap} in the 12--28~GeV/c
interval to $\sim$19\%~\cite{detpap} at about 70~\GeVc, as measured
with test-beam muons.  
Given the relatively small
size of the energy bins, the average neutrino energy is very similar
for charm production events and CC events within the same bin, and no
correction is necessary.  
The efficiency is calculated by weighting
the energy-dependent and decay topology-dependent efficiencies with
the measured branching ratios as reported above.

\begin{figure}[tb]
  \begin{center} 
   \scalebox{0.45}{\includegraphics{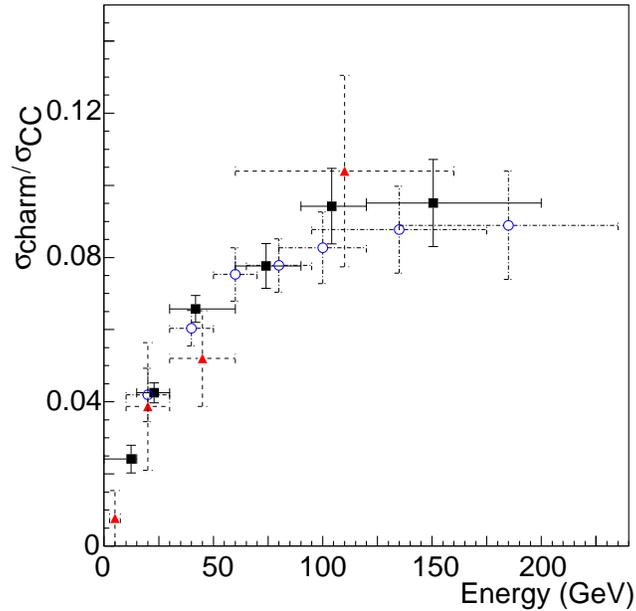}}
  \caption{ Energy dependence of the relative inclusive charm production 
  cross-section ratio. The squares show the
  measurements reported here, the points marked with triangles the
  E531 result. The circles represent the dimuon cross-section measured
  in Ref.~\cite{CHORUS_dimuon} scaled for the muonic branching ratio quoted in
  this paper. \label{fig:energy}
  }
\end{center}
\end{figure}

The energy dependence of backgrounds is assumed to be the same as the one of
charged current neutrino events.
The differential cross-section measurement is normalized to the total
neutrino--nucleon cross-section and thus is not affected by the
uncertainties between the beam simulation and the beam flux
measurement.
The measurement of the charm production rate relative to the CC
interaction rate is shown as function of neutrino energy and compared
with the measurement from E531~\cite{E531} in Figure~\ref{fig:energy}.
Good agreement with an improved precision with respect to E531
measurement is shown.  
Very good agreement is found with respect to the
dimuon cross-section measured with the CHORUS electronic detector by scaling
the dimuon results for the muonic charm decay fraction quoted in this
paper.

\begin{figure}[tb]
  \begin{center} 
   \scalebox{0.45}{\includegraphics{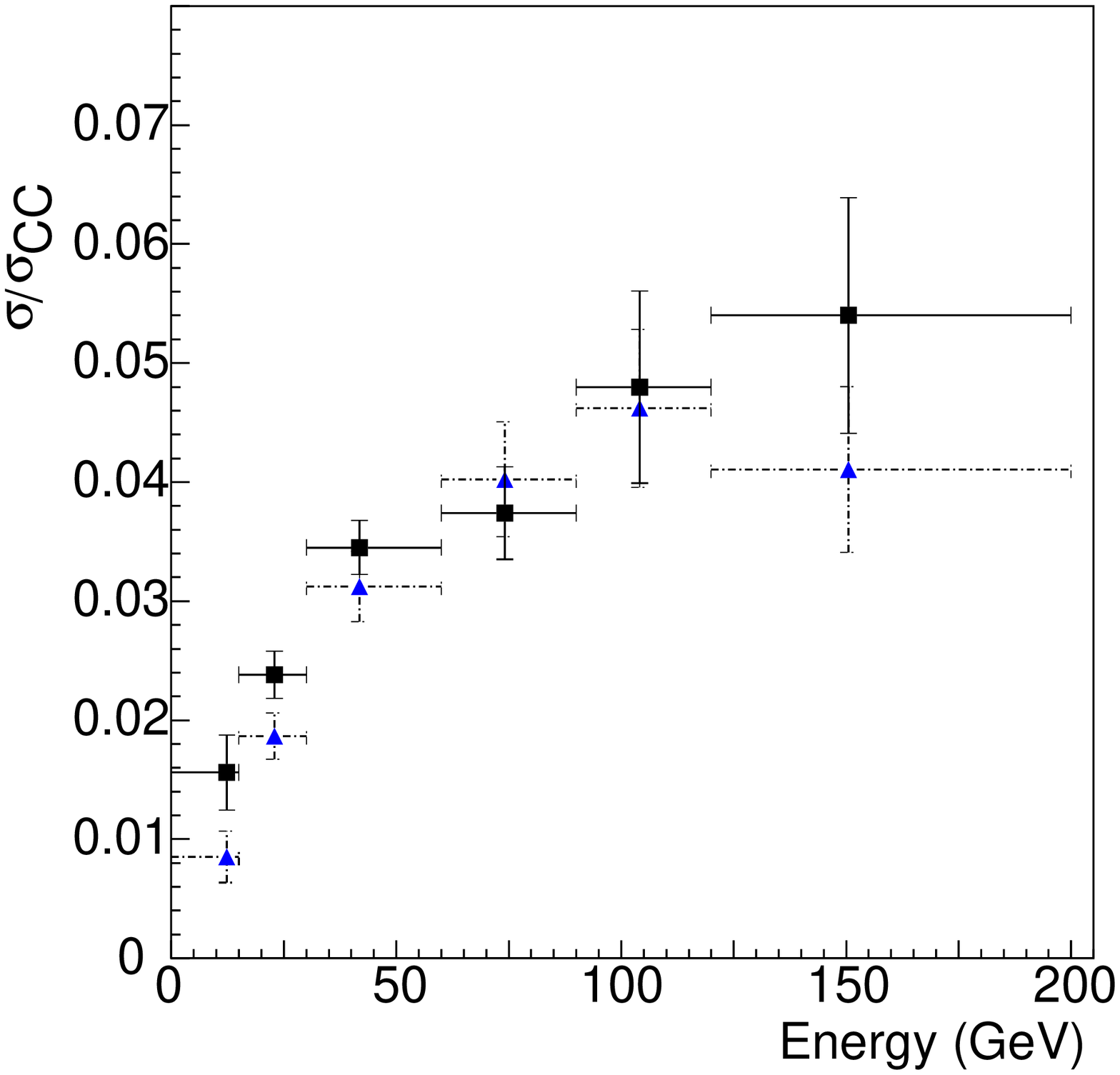}}
  \caption{ Energy dependence for charged (squares) and neutral
  (triangles) charm cross-section ratio relative to the CC cross-section.
  \label{fig:energy2}
  }
\end{center}
\end{figure}

The energy dependence for charged and neutral charm is reported
separately in Figure~\ref{fig:energy2}. A very similar energy behaviour
is shown except for the low energy region where the contribution of
quasi-elastic production of \lbcpl may account for the difference~\cite{QE}.

\section{Acknowledgements}

We gratefully acknowledge the help and support of the neutrino-beam
staff and of the numerous technical collaborators who contributed to
the detector construction, operation, emulsion pouring, development,
and scanning. The experiment was made possible by grants from the
Institut Interuniversitaire des Sciences Nucl$\acute{\mathrm{e}}$aires
and the Interuniversitair Instituut voor Kernwetenschappen (Belgium);
the Israel Science Foundation (Grant 328/94) and the Technion Vice
President Fund for the Promotion of Research (Israel); CERN (Geneva,
Switzerland); the German Bundesministerium f$\ddot{\mathrm{u}}$r
Bildung und Forschung (Germany); the Institute of Theoretical and
Experimental Physics (Moscow, Russia); the Istituto Nazionale di
Fisica Nucleare (Italy); the Promotion and Mutual Aid Corporation for
Private Schools of Japan and Japan Society for the Promotion of
Science (Japan); Korea Research Foundation Grant (KRF-2003-005-C00014)
(Republic of Korea); the Foundation for Fundamental Research on Matter
FOM and the National Scientific Research Organization NWO (The
Netherlands); and the Scientific, and Technical Research Council of
Turkey (Turkey).  We gratefully acknowledge their support.


\begin{thebibliography}{99}

\bibitem{jpsislac} J. E. Augustin {\it et al.}, {\it Phys. Rev. Lett.} {\bf 33} (1974) 1406.

\bibitem{jpsibnl} J. J. Aubert {\it et al.}, {\it Phys. Rev. Lett.} {\bf 33} (1974) 1404.

\bibitem{niucharm} 
	K.~Niu, E.~Mikumo, Y.~Maeda,
	{\it  Prog. Theor. Phys.} {\bf 46} (1971) 1644.

\bibitem {opera} 
	 R. Acquafredda {\it et al.}, {\it JINST} {\bf 4}:P04018 (2009).

\bibitem {nufact} J.J. Gomez-Cadenas {\it et al.}, 
	 "Physics opportunities at Neutrino factories", 
	 {\it Ann. Rev. Nucl. Part. Sci.} {\bf 52} (2002) 253.

\bibitem{CDHS}H. Abramowicz {\it et al.}, CDHS Collaboration, {\it  Z. Phys.} {\bf C15} (1982) 19.

\bibitem{CCFR}S.A. Rabinowitz {\it et al.}, CCFR Collaboration,  {\it  Phys. Rev. Lett. } 
{\bf 70} (1993) 134.

\bibitem{CHARM}M. Jonker {\it et al.},  CHARM Collaboration, {\it Phys. Lett.} {\bf B107} (1981) 241.

\bibitem{CHARMII}P. Vilain {\it et al.}, CHARM II Collaboration,
{\it Eur. Phys. J.} {\bf C11} (1999) 19.

\bibitem{NuTeV}M. Goncharov {\it et al.}, NuTeV Collaboration, {\it Phys. Rev.} {\bf D64} (2001) 112006.

\bibitem{CHORUS_dimuon} A. Kayis-Topaksu {\it et al.}, CHORUS Collaboration, {\it Nucl. Phys.} {\bf B798} (2008) 1.

\bibitem{BEBC}
    A. E. Asratian {\it et al}., {\it Z. Phys.} {\bf C68} (1995) 43.

\bibitem{NOMAD}P. Astier {\it et al.},  NOMAD Collaboration, {\it Phys. Lett. } {\bf B486} (2000) 35.

\bibitem{dstar}  G. Onengut {\it et al.}, CHORUS Collaboration, {\it Phys. Lett.} {\bf B614} (2005) 155.

\bibitem{E531}N. Ushida, {\it et al.}, E531 Collaboration, {\it Phys. Lett. } 
{\bf B206} (1988) 375.

\bibitem{detpap}E. Eskut, {\it et al.}, CHORUS Collaboration, {\em Nucl.
Instrum. and Methods} {\bf A401} (1997) 7.

\bibitem{aoki}S. Aoki  {\it et al}., {\it Nucl. Instrum. and Methods} {\bf B51}
  (1990) 466.

\bibitem{TrackSelector}T. Nakano. Ph.D. thesis, Nagoya University, Japan, 1997.

\bibitem {beam} E. H. M. Heijne, CERN Yellow Report 83-06 (1983).

\bibitem{dzero}  G. Onengut {\it et al.}, CHORUS Collaboration, {\it Phys. Lett.} {\bf B613} (2005) 105.

\bibitem{det_trig}M.G. van Beuzekom, {\it et al.}, CHORUS Collaboration, {\em Nucl.
Instrum. and Methods} {\bf A427} (1999) 587.

\bibitem{murat}A. M. G{\" u}ler, Ph.D. thesis, Middle East Technical University, Ankara, Turkey (2000).

\bibitem{bart}B. Van de Vyver, Ph.D. thesis, Vrije Universiteit Brussel, Brussels, Belgium, 2002. CERN-THESIS-2002-024.

\bibitem{NetScan}N. Nonaka, Ph.D. thesis, Nagoya University, Japan (2002).

\bibitem{ilyaproceedings} I. Tsukerman, CHORUS Collaboration, MC
Generators in CHORUS, {\it Nucl. Phys. Proc. Suppl.} {\bf 112} (2002)
177.

\bibitem{gbeam}S. Sorrentino, Diploma Thesis, Naples University, Italy (1995).

\bibitem{geant}GEANT 3.21, CERN program library long write up W5013.

\bibitem{fluka98}A. Fass\`{o} {\it et al.}, SARE-3 Workshop, KEK Report Proceedings
    97-5 (1997) 32.

\bibitem{jetta} P.~Zucchelli, Ph.D. thesis, Universit\'a di Ferrara, 
Italy (1995).

\bibitem{lepto}G. Ingelman, {\it Preprint TSL/ISV} 92-0065, Uppsala University, Sweden (1992).

\bibitem{jetset}T. Sj\"ostrand, {\it Comput. Phys. Commun.} {\bf 82} (1994) 74.

\bibitem{resque}S. Ricciardi, Ph.D. thesis, Universit\'a di Ferrara, 
Italy (1996).

\bibitem{qegen}F. Di Capua, Ph.D. thesis, Universit\'a di Napoli,
Italy (2003).

\bibitem{astra} O. Melzer, Diploma thesis, Westf\"{a}lische
    Wilhelms-Universit\"{a}t, M\"{u}nster, Germany (1997).

\bibitem{QE}A. Kayis-Topaksu {\it et al.}, CHORUS Collaboration, {\it
  Phys. Lett.} {\bf B575} (2003) 198.

\bibitem{delellis}G. De Lellis {\it et al.}, {\it Phys. Lett.} {\bf B550} (2002) 16.

\bibitem{Satta}A. Satta, Ph.D. thesis, Universit\'a La Sapienza
di Roma, Italy (2001).

\bibitem{PetreraRomano} S. Petrera and G. Romano, {\it
Nucl. Instrum. and Meth} {\bf 174} (1980) 61.

\bibitem{Bolton2} T. Bolton, arXiv:hep-ex/9708014. 

\bibitem{PDG} Particle Data Group, {\it J. of Physics} G{\bf 37} (2010) 075021.

\bibitem{lambdac}A. Kayis-Topaksu {\it et al.}, CHORUS Collaboration, {\it Phys. Lett.} {\bf B555} (2003) 156.

\bibitem{Bmu2} A. Kayis-Topaksu {\it et al.}, CHORUS Collaboration, {\it
Phys. Lett.} {\bf B626} (2005) 24.

\bibitem{AntiCharm}  G. Onengut {\it et al.}, CHORUS Collaboration, {\it
Phys. Lett.} {\bf B604} (2004) 145.


\end{thebibliography}
\end{document}